\documentclass[twocolumn,showpacs,amsmath,amssymb,prb]{revtex4}

\usepackage{graphicx}% Include figure files
\usepackage{bm}% bold math
\begin{document}

\title{Magnetization reversal of ferromagnetic nanodisk placed above a superconductor}
\author{A. A. Fraerman$^{1,2}$ }
\author{I. R. Karetnikova }
\author{I. M. Nefedov }
\author{I. A. Shereshevskii }
\author{M. A. Silaev }
\affiliation{$^1$ Institute for Physics of Microstructures, Russian
Academy of Sciences, 603950, Nizhny Novgorod, GSP-105,
 Russia,\\
 $^2$ Argonne National Laboratory, Argonne, Illinois, 60439}

\date{\today}
\begin{abstract}
Using numerical simulation we have studied a magnetization
distribution and a process of magnetization reversal in nanoscale
magnets placed above a superconductor plane. In order to consider
an influence of superconductor on magnetization distribution in
the nanomagnet we have used London approximation. We have found that for
usual values of London penetration depth the ground state
magnetization is mostly unchanged. But at the same time the fields
of vortex nucleation and annihilation change significantly: the
interval where vortex is stable enlarges on 100-200 Oe for the
particle above the superconductor. Such fields are experimentally
observable so there is a possibility of some practical
applications of this effect.
\end{abstract}


\maketitle

\section{Introduction} \label{sec:intro}

 In the past few years considerable attention
has been devoted to the investigations of magnetism in small
nano-sized ferromagnetic particles. Such interest is caused by
the opportunities for creating recording devices\cite{Novosad1}$^-$\cite{Novosad3}
and ultrasmall magnetic field sensors\cite{Novosad4} based on the
properties of ferromagnetic particles. It is now well-understood
that magnetization distribution in a single particle is determined by
the competition between the magnetostatic and exchange energies.
If a particle is small, it is uniformly magnetized and if
its size is large enough a non-uniform(vortex) magnetization is more
energy preferable(see, for example, Refs. \onlinecite{Novosad4},
\onlinecite{From_Frayerman_JMMM1}, \onlinecite{From_Frayerman_JMMM2}, \onlinecite{Novosad5}).
Besides the geometrical form and size, the state of the
particle depends on many other factors. For example, by applying
a homogeneous magnetic field we can cause nucleation or
annihilation of the vortex. In an array of particles their magnetostatic
interaction may have a strong influence on a particle
magnetization. If the distance between particles
is rather small then magnetostatic interaction has
a strong destabilizing effect on the vortex state, leading to a
significant decrease in both the nucleation and annihilation
fields \cite{Guslienko}$^,$\cite{Frayerman_phr}.

The interplay between the
ferromagnetism and superconductivity can also lead to changes in the
magnetization distribution. It was shown that in a ferromagnetic
film put on a superconducting substrate the size of the domains is up to
$\sqrt{1.5}$ times smaller than for a film without a superconducting
substrate\cite{Sonin}$^,$\cite{Bulaevscij}. The experimental
investigations revealed changes in the magnetic field around Al/Ni
submicron structures with a decrease in the temperature to values below $T_c$. This effect
was referred to the expulsion of the magnetic field by the
superconducting part of the Ferromagnetic/Superconductor (FS)
hybrid structures\cite{Dubonos}.

All investigations of the FS interaction deal with the changes of the ferromagnetic
domain structure. Therefore the results thereof cannot be
applied to the single-domain nanoparticles. On the one hand, the magnetization
of a nanoparticle is more simple than the domain structure of a macroscopic
ferromagnetic, therefore, theoretical findings could be proved
by experiments with nanoparticles. But on the other hand, the
magnitude of the interaction between the nanoparticles and the
superconductor is strongly reduced due to the large values of the London penetration
depth.

In this work we investigate the phase transition between the single-domain
and the vortex state, and the process
of magnetization reversal of a ferromagnetic nano-sized particle
placed above the surface of a superconductor. The
superconducting state is described using the London approximation, that
is, we assume that the particle cannot produce a vortex- antivortex
pair. This assumption is not universally true, since if the particle dimensions are
sufficiently large its magnetic field can
destroy the Meijssner state of the superconductor\cite{Mochalkov}. The
criteria of the London approximation applicability can be obtained in the following way.
The largest magnetic field is generated by a
single-domain particle. Then, the maximal magnetic flux through the
surface of a superconductor is $\phi_m=\pi R h M$, where
$R$ and $h$ are the radius and the height of the particle and M is
the magnetization. The flux $\phi_m$ must be less than the flux
quantum $\phi_0=2,07\cdot 10^{-7} gs\cdot sm^2$ in order to
exclude the possibility of a vortex penetration. We consider a
ferromagnetic particle with a saturation magnetization of about $M_s=800\, oe$. Then we
have the following limitation on the particle dimensions:
$Rh<8\cdot 10^3\, nm^2$. So we consider only the particles
which obey this condition.

 The equilibrium distribution of magnetization which gives a minimum
to the energy functional is found by numerical simulation. For
numerical simulation we use the approach based on the
Landau-Lifshitz-Gilbert(LLG) equation for the dynamics of magnetic
moments. This approach enables us to investigate metastable states
which realize the local minimums of the energy functional. The
metastable states are of great importance since the finiteness
of the nucleation and annihilation fields values is the consequence of the
energy barrier between single-domain and vortex-like states. We
obtain a phase diagram of the particle in the height/diameter plane
for the transition from the single-domain  to the vortex-like
state. We find that for the realistic value of the London penetration depth
the FS interaction is rather weak. The energy of the FS interaction is
about 100 times smaller than the self energy of the particle.
Therefore, the FS interaction has no influence on the phase diagram.
Nevertheless, the magnetization curve of the particle in an external
homogeneous magnetic field changes significantly. Even the 1/100
energy shift means that the superconducting screening current
produces a magnetic field of about $4\pi M/100\sim 100 Oe$ which
leads to experimentally observable decrease of the vortex
nucleation field and increase of the annihilation field. The paper is
organized as follows. In Sec.2 we derive the analytical expression
for the energy functional of our system. In Sec.3 we show the most
important features of the numerical simulation. In Sec.4 we
discuss the results and also propose possible experiments.
 Finally, the summary is given in Sec.5.

\section{Energy functional} \label{sec:energy}

    \begin{figure}[t]
    \centerline{\includegraphics[width=0.7\linewidth]{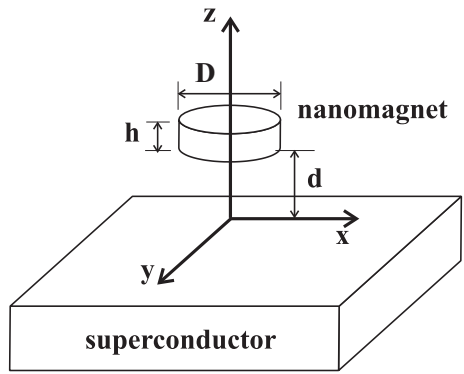}}
 %\begin{center}
 %\epsfxsize=70mm \epsfbox{}
 %\end{center}
 \caption{Ferromagnetic nanoparticle above the surface of superconductor.}
 \label{particle}
 \end{figure}

 The system we consider is a ferromagnetic particle placed above
 the surface of a superconductor
 (Fig.\ref{particle}). We assume that the superconductor
 occupies the whole half-space, so it has only one boundary plane, $z=0$.
 The ferromagnetic particle is assumed to be made of soft magnetic material,
 so the energy of anisotrophy is equal to zero. The total energy of
 the system of particle and
  superconductor reads:

 \begin{equation}\label{energy0}
 E=E_{e}+E_m+E_{ext}.
 \end{equation}
 The first term $E_{e}$ is the energy
 of the exchange interaction:

 \begin{equation}\label{exch_energy}
 E_{e}=-\frac{J}{2M_s^2}\int_V \left(|\nabla M_x|^2+|\nabla M_y|^2+|\nabla M_z|^2\right)
  d^3{\bf{r}},
 \end{equation}
 where $J$ is the constant of exchange interaction.
 The second term $E_{m}$ in Eq.(\ref{energy0}) is  the magnetostatic energy\cite{Pokrovsky}:

 \begin{equation}\label{magn_energy}
 E_m=-\frac{1}{2}\int_V{\bf{H}}\cdot{\bf{M}}\,d^3{\bf{r}}.
 \end{equation}
 Here ${\bf{H}}$ is the sum of magnetic field ${\bf{H}}_m$ induced by the particle and
 magnetic field ${\bf{H}}_s$ generated by the superconducting current:
 $$
 {\bf{H}}={\bf{H}}_m+{\bf{H}}_s.
 $$
 The last term $E_{ext}$ in Eq.(\ref{energy0}) is the energy of the interaction between the magnetic moment
 and the external field ${\bf{H}}_0$:
 \begin{equation}\label{ext_energy}
 E_{ext}=-\int_V  {\bf{H}}_0 \cdot {\bf{M}} d^3{\bf{r}}.
 \end{equation}

 In order to obtain the energy
 as the functional of ${\bf{M}}$ we should express the magnetic field
 ${\bf{H}}$ in terms of ${\bf{M}}$. To this end we should solve the
 Maxwell-London equations where the source is a magnetic current
 ${\bf{j}}_m=c\nabla\times{\bf{M}}$.

 In the space outside the superconductor the system is described by the
 following equations:

  \begin{equation}\label{Maxwell1}
  \nabla\times{\bf{B}}=\frac{4\pi}{c}{\bf{j}}_m
  \end{equation}

  \begin{equation}\label{Maxwell1.0}
  \nabla\cdot{\bf{B}}=0
  \end{equation}
  where we introduce the vector of magnetic induction ${\bf{B}}={\bf{H}}+4\pi{\bf{M}}$.
  The superconductor is described by the equations:

 \begin{equation}\label{Maxwell2}
 \nabla\times{\bf{B}}=\frac{4\pi}{c}{\bf{j}}_s
 \end{equation}
 \begin{equation}\label{London}
 {\bf{B}}=-\lambda_L^2\frac{4\pi}{c}\nabla\times{\bf{j}}_s,
 \end{equation}
 where ${\bf{j}}_s$ is the superconducting current and $\lambda_L$
 is the London penetration depth. For the boundary conditions we take the
 continuity of the magnetic induction ${\bf{B}}$ and its derivatives at
 the boundary plane $z=0$.
 Since the Maxwell-London equations are linear, we first can consider
 a single magnetic dipole placed at the point
 ${\bf{r}}_0=(x_0,y_0,z_0)$ above the surface of the superconductor:
 $$
 {\bf{M}}={\bf{m}}\delta({\bf{r}}-{\bf{r}}_0),
 $$
 where ${\bf{m}}$ is the magnetic dipole moment.
 Once this problem is solved, we can construct the solution for any
 magnetization distribution. After a bit troublesome but
 straightforward computation, we obtain the solution to the
 single-dipole problem at the half-space $z>0$ in the form:
 $$
 {\bf{B}}=-\nabla\phi+4\pi{\bf{M}},
 $$
 where the magnetic potential $\phi$ is:

\begin{equation}\label{s_pot}
  \phi=\phi_d+\phi_{refl}+\phi_{\lambda}.
\end{equation}
The first term $\phi_d$ in Eq.(\ref{s_pot}) is the potential of a single
magnetic dipole without a superconductor:
  $$
 \phi_d({\bf{r}})=-{\bf{m}}\cdot \nabla \frac{1}{|{\bf{r}}-{\bf{r}}_0|},
 $$
The second term $\phi_{refl}$ is the potential of the magnetic dipole
reflected at the plane $z=0$:
  $$
 \phi_{refl}({\bf{r}})=-{\bf{m}}^*\cdot \nabla \frac{1}{|{\bf{r}}-{\bf{r}}_0^*|},
  $$
  where
  $$
  {\bf{r}}_0^*=(x_0, y_0,-z_0),
  $$
  $$
  {\bf{m}}^*=(m_x, m_y, -m_z).
  $$
   The third term $\phi_{\lambda}$ in Eq.(\ref{s_pot}) appears due to the finiteness of
  the London penetration depth value $\lambda_L$. The expression for
  $\phi_{\lambda}$ is:

\begin{equation}\label{potential_lambda}
\phi_{\lambda}({\bf{r}})=-{\bf{m}}^*\cdot \nabla
g({\bf{r}}-{\bf{r}}_0^*),
\end{equation}
 where
 $$
 g({\bf{r}})=\lambda_L^2\int_{0}^{\infty}{2k^2\exp(-kz)J_0(k\rho)dk}-
 $$
 $$
 \lambda_L^2\int_{0}^{\infty}{2k(k^2+\lambda_L^{-2})^{1/2}\exp(-kz)J_0(k\rho)dk},
 $$
 $$
 \rho=(x^2+y^2)^{1/2}.
 $$
 Here $J_0(x)$ is the Bessel function.
 Then the single dipole magnetic field above the
 surface of the superconductor is expressed as:

 $$
 {\bf{B}}({\bf{r}})={\bf{H}}({\bf{r}})+
 4\pi{\bf{m}}\delta({\bf{r}}-{\bf{r}}_0),
 $$
 $$
  {\bf{H}}({\bf{r}})=\widehat{{\bf{D}}}({\bf{r}},{\bf{r}}_0){\bf{m}}
 $$
 where $\widehat{{\bf{D}}}$ is the modified dipole matrix:

\begin{equation}\label{matr}
\widehat{{\bf{D}}}({\bf{r}},{\bf{r}}_0)=
 \widehat{{\bf{D}}}_d({\bf{r}}-{\bf{r}}_0)
 + \widehat{{\bf{D}}}_{s}({\bf{r}}-{\bf{r}}_0^*),
  \end{equation}
 Here the first term is an ordinary dipole matrix:
 $$
  D_{d x_1x_2}({\bf{r}})=\frac{\partial^2}{\partial x_1\partial x_2}\frac{1}{|{\bf{r}}|},
 $$
 and the second term appears due to the presence of the superconductor

  $$
 D_{s x_1x_2}({\bf{r}})=(-1)^{\delta_{z,x_2}}
 \frac{\partial^2}{\partial x_1\partial x_2} {\left(g({\bf{r}})+\frac{1}{|{\bf{r}}|}\right)}.
 $$
 where indexes $x_1$ and $x_2$ denote coordinates $x,y,z$ and
 $\delta_{z,x_2}$ is the Kroneker's symbol.

 Let us now consider continuous distribution of the magnetic
 moment ${\bf{M}}({\bf{r}})$. The magnetic field generated by the
 superconducting current is given as a sum of single-dipole
 solutions:

 $$
 {\bf{H}}_s=\int_V
 \widehat{{\bf{D}}}_{s}({\bf{r}}-{\bf{r}}_0^*){\bf{M}}({\bf{r}}_0)
 d^3{\bf{r}}_0.
 $$
 For the magnetic field ${\bf{H}}_m$ generated by the magnetic current we cannot
 take the sum of the single-dipole fields since the dipole matrix
 $\widehat{{\bf{D}}}_d({\bf{r}})$ has a nonintegrable singularity
 at point ${\bf{r}}=0$. Therefore, the expression for
 ${\bf{H}}_m$ that we use is:

 $$
 {\bf{H}}_m({\bf{r}})=-\int_V
 {\nabla\cdot{\bf{M}}({\bf{r}}_0)}\frac{{\bf{r}}-{\bf{r}}_0}{|{\bf{r}}-{\bf{r}}_0|^3}
 d^3{\bf{r}}_0+
 $$
 $$
 \int_{\partial V}
 ({{\bf{M}}({\bf{r}}_0)}\cdot{{\bf{n}}}_s({{\bf{r}}}_0))\frac{{\bf{r}}-{\bf{r}}_0}{|{\bf{r}}-{\bf{r}}_0|^3}
 d^3{\bf{r}}_0.
 $$
 Here, $V$, $\partial V$ are the volume and surface of a particle
 and ${{\bf{n}}}_s$ is the unit vector of the external normal to the
 surface at a current point.

 Finally, we consider the energy functional consisting of the self-energy of particle $E_0$,
 the energy of interaction with the superconductor $E_{int}$ and with external field $E_{ext}$:
 $$
 E=E_0+E_{int}+E_{ext}.
 $$
 Self-energy $E_0$ includes exchange $E_e$ energy and the part $E_d$ of magnetostatic energy $E_m$:

 $$
 E_0=E_e+E_d,
 $$
 where
 $$
 E_d=-\frac{1}{2}\int_V {\bf{H}}_m\cdot{\bf{M}}\, d^3{\bf{r}}=
 $$

 $$
 -\frac{1}{2}\int_{V\times V}
 (\nabla\cdot{\bf{M}}({\bf{r}}))(\nabla\cdot
 {\bf{M}}({\bf{r}}'))
 \frac{d^3{\bf{r}}d^3{\bf{r}}'}{|{\bf{r}}-{\bf{r}}'|}
 $$
 $$
 + \frac{1}{2}\int_{V\times \partial V}
 (\nabla\cdot {\bf{M}}({\bf{r}}'))({\bf{M}}({\bf{r}})\cdot
 {\bf{n}}_s)
 \frac{d^3{\bf{r}}d^3{\bf{r}}'}{|{\bf{r}}-{\bf{r}}'|}
 $$
 $$
 -\frac{1}{2}\int_{\partial V\times \partial V}
 ({\bf{M}}({\bf{r}}')\cdot {\bf{n}}'_s)({\bf{M}}({\bf{r}})\cdot
 {\bf{n}}_s)
 \frac{d^3{\bf{r}}d^3{\bf{r}}'}{|{\bf{r}}-{\bf{r}}'|}.
 $$
 The energy of interaction $E_{int}$ of the particle with the superconductor is the energy
 of the magnetic moment in the field produced by the superconducting
 current, and it is the other part of
 the magnetostatic energy $E_m$:

 $$
 E_{int}=-\frac{1}{2}\int_V \,{\bf{H}}_s\cdot {\bf{M}}\,d^3{\bf{r}}=
 $$
 $$
 -\frac{1}{2}\int_{V\times V} \,
  {\bf{M}}({\bf{r}})\cdot
  \widehat{{\bf{D}}}_s({\bf{r}},{\bf{r}}_0){\bf{M}}({\bf{r}}_0)\,
  d^3{\bf{r}}\,d^3{\bf{r}}_0
 $$
 The energy $E_{ext}$ is the same as in Eq.(\ref{energy0}) and is given by
 Eq.(\ref{ext_energy}).

  \section{Numerical simulation} \label{sec:numerical}

  To perform numerical calculation we use the same approach as in
  Refs. \onlinecite{Frayerman_phr} and \onlinecite{Fraerman_metalls}.
  The basis of our numerical simulation is the LLG equation for
  magnetization {\bf{M}}({\bf{r}},t) of a particle in the form
 \begin{equation}\label{LLG}
  \frac{\partial {\bf{M}}}{\partial t}=
  -\frac{\gamma}{1+\alpha^2}[{\bf{M}},{\bf{H_{eff}}}]-
  \frac{\alpha\gamma}{(1+\alpha^2)M_s}[{\bf{M}},[{\bf{M}},{\bf{H}}_{eff}]],
 \end{equation}
 where $\gamma$ is the gyromagnetic ratio, $\alpha$ is the
 dimensionless damping parameter and $t$ is time.
 The effective field ${\bf{H}}_{eff}$ is a variation derivative of the energy functional:

 $$
 {\bf{H}}_{eff}=-\frac{\delta E}{\delta {\bf{M}}}
 $$

 The important feature of the LLG equation is that it describes
 the evolution of the magnetization distribution to the
 equilibrium. By varying the initial
 conditions we find different equilibrium states of our system. Then we
 choose the equilibrium state with the minimal energy to obtain the ground
 state. Choosing as the initial state the vortex-like or single-domain
 distribution and varying the external field ${\bf{H}}_0$ we determine the
 vortex annihilation or nucleation field as the critical field
 of transformation of the vortex-like to single-domain state or vice versa.

 To avoid a three-dimensional grid problem to be solved which needs large computer resources
 we assume that magnetization of a cylindrical particle does not depend on the
 coordinate $z$ along the cylindrical axis. Then we integrate the
 relations for the energy over $z$ and $z_0$ and obtain the energy
 as a functional of the magnetization which is a function of only two
 space variables. Then we define the effective field ${\bf{H}}_{eff}$ as a variation
 derivative of the obtained functional. The effective field does not depend on
 $z$ either, so we have a three-dimensional problem reduced to the two-dimensional one.

 To develop a numerical method we divide the particle into rectangular parallelepipeds
 with a square base of size $a$ in the plane $(x,y)$ and of height $h$ and
 obtain approximate expressions for different parts of the energy
 functional using the grid values of magnetization
 ${\bf{M}}=(M_x({\bf{\rho}}), M_y({\bf{\rho}}),
 M_z({\bf{\rho}}))$.

The expression for the magnetostatic energy $E_d$ reads:

\begin{equation}\label{approximate_magnetostatic energy}
  E_d[{\bf{M}}]=-\frac{a^4}{2}\sum_{\rho\neq{\rho}'}
  {\bf{M}}({\rho}')\cdot\hat{\bf{D}}_d^h(\rho-{\rho}'){\bf{M}}(\rho)+E_d^0,
\end{equation}
  where ${\bf{\rho}}=(x,y)$ are the points of the square grid with
  step $a$ on the $(x,y)$ plane and matrix $\hat{\bf{D}}_d^h$ is
  the dipole matrix $\hat{\bf{D}}_d$ integrated over the $z$
  coordinate:
  $$
 \hat{\bf{D}}_d^h({\bf{\rho}})=
 \int_{0}^{h}{dz\int_{-z}^{h-z}{\hat{\bf{D}}_d({\bf{\rho}},{z}')\,d{z}'}}.
 $$
 The additional term $E_d^0$ in
 Eq.(\ref{approximate_magnetostatic energy}) appears as the
 contribution of self-interaction in the cell. It depends only on the value of magnetization
 which is assumed constant, so it does not influence the effective field.

The expression for the exchange energy reads:
 \begin{equation}\label{approximate_exchange energy}
 E_e[{\bf{M}}]=-\frac{Jh}{2M_s^2}\sum_{\rho}\sum_{{\rho}'}
 |{\bf{M}}({\bf{\rho}})-{\bf{M}}({\bf{\rho}}')|^2.
 \end{equation}
 The internal summation in Eq.(\ref{approximate_exchange energy}) is
 taken over all neighbors ${\bf{\rho}}'$ of the point
 ${\bf{\rho}}$.

 The energy $E_{ext}$ and the interaction energy $E_{int}$ are,
 respectively:
 \begin{equation}\label{approximate_external field energy}
 E_{ext}=-a^2h\sum_{\rho}{\bf{H}}_0\cdot{\bf{M}}({\bf{\rho}});
 \end{equation}

 \begin{equation}\label{approximate_interaction energy}
  E_{int}[{\bf{M}}]=-\frac{a^4}{2}\sum_{\rho\neq{\rho}'}
  {\bf{M}}({\bf{\rho}}')\cdot
  \hat{\bf{D}}_s^h({\bf{\rho}}-{\bf{\rho}}')
  {\bf{M}}({\bf{\rho}}),
  \end{equation}
 where matrix $\hat{\bf{D}}_s^h$ is

 $$
 \hat{\bf{D}}_s^h({\bf{\rho}})=
 \int_{d}^{d+h} dz \int_{z+d}^{z+d+h} \hat{\bf{D}}_s({\bf{\rho}},{z}')\,d{z}'.
 $$

 We choose the cell size considering two factors.
 On one hand the size of the cell should be
 smaller than the characteristic exchange interaction length ($\sqrt{J/M_s^2}$) in
 order to describe the inhomogeneous magnetization correctly. On
 the other hand, we cannot choose it very small because of the
 computation time limitations.

  \section{Results and discussion} \label{sec:results}

  For computation we choose the following parameters: the saturation
  magnetization $M_s=800 Oe$, the exchange interaction constant $J=10^{-6}
  erg/sm$. The cell size is $3nm\times 3nm$, while the exchange
  interaction length is approximately $13 nm$.
  The distance $d$ and the London penetration depth have a strong
  influence on the interaction between a particle and a
  superconductor. For superconductors with $\lambda_L\sim 50-100 nm$
  the interaction should be largely reduced, since such values of $\lambda_L$ are
  comparable to the size of the particle. To compensate for this reduction we can make the distance
  $d$ small. We choose $\lambda=50 nm$ and $d=5 nm$ since such
  values are experimentally obtainable and make the effect of a
  particle-superconductor interaction quite distinctive.

\begin{figure}[h]
   \centerline{\includegraphics[width=0.7\linewidth]{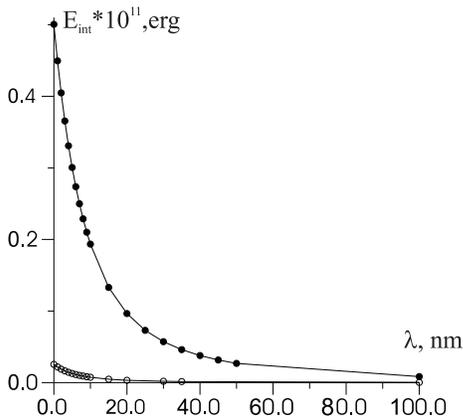}}
%\begin{center}
%\epsfxsize=50mm \epsfbox{.eps}
%\end{center}
\caption{ The energy of interaction of a single-domain (filled
circles) and vortex-like (open circles) magnetized particle
with a superconductor as a function of $\lambda_L$.}
\label{energy}
\end{figure}

 First of all, we investigate how the energy of the interaction between
 the particle and the superconductor depends on the value of
 $\lambda_L$. We calculate the interaction energy $E_{int}$ for the particle with a single-domain
 and vortex-like magnetization(Fig.\ref{energy}). The dimensions of the particle are: $h=10
 nm$ and $D=50 nm$. The self-energy of this particle is $E_{0}=1.2122\cdot
 10^{-11} erg$ for the single domain state and $E_0=1.8856\cdot 10^{-11}
 erg$ for the vortex one. The interaction energy decreases rapidly when $\lambda_L$
 changes from $0$ to $30-40 nm$. For larger values of $\lambda_L$
 the interaction energy tends to zero asymptotically. At
 practically interesting values $\lambda_L\sim 50-100 nm$ the
 interaction energy is about $E_{int}\sim 0.0268\cdot 10^{-11}
 erg$ for the single-domain magnetization and $E_{int}\sim 0.0014
\cdot 10^{-11} erg$ for the vortex-like magnetization. Thus the
interaction with a superconductor is much stronger for a
single-domain particle. This is easily understood because the vortex-like
magnetization produces a much weaker magnetic field than the
single-domain magnetization. The interaction energy for the real
values of $\lambda_L\sim 50 nm$ is much smaller than the particle
self energy: for the single-domain magnetization it is
$E_{int}/E_{0}\sim 0.02$.

\begin{figure}[h]
   \centerline{\includegraphics[width=0.7\linewidth]{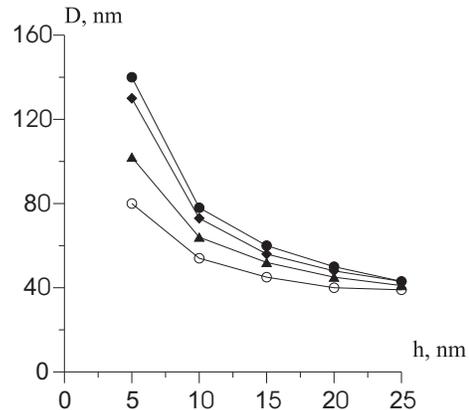}}
%\begin{center}
%\epsfxsize=60mm \epsfbox{.eps}
%\end{center}
 \caption{ Phase diagram of the vortex-single domain state
 transition. Filled circles- $\lambda_L=\infty$,
 open circles- $\lambda_L=0$, triangles- $\lambda_L=10 nm$,
 rhombus- $\lambda_L=50 nm$. }
 \label{diagram}
 \end{figure}

Then we study the ground state of the particle interacting with
the superconductor. We define the ground state as a stable state
with the lowest energy. There are two possible stable states for
the particle: vortex-like and single-domain states. We
obtain a phase diagram (see Fig.\ref{diagram}) where in the area
above the phase boundary the ground state is vortex-like and
below the boundary it is single-domain. Different curves on
Fig.\ref{diagram} correspond to different values of $\lambda_L$.

Each phase boundary on Fig.\ref{diagram} has basic
properties coming from the origin of phase transitions in
a ferromagnetic nanoparticle\cite{Cowburn}. They are determined by
the interplay between the magnetostatic and exchange energy. In
relatively large particles the ground state is vortex-like.
When the dimensions(diameter $D$ or height $h$) of the particle
are reduced then the single domain state becomes the ground one.
The superconductor influence shows up through an enlargement of the area of
small-sized particles having the vortex-like state.
It is clear that the shift of the phase boundary in
comparison to the case $\lambda=\infty$ depends on the ratio of
the interaction energy to the self-energy of the single-domain
particle: $E_{int}/E_0$. Since this ratio is very small
$E_{int}/E_0=0.02\ll 1$ at $\lambda_l=50 nm$, the phase
boundary remains almost unchanged.

\begin{figure}[h]
   \centerline{\includegraphics[width=0.7\linewidth]{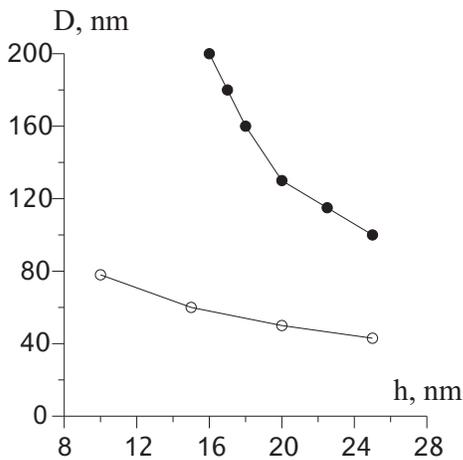}}
%\begin{center}
%\epsfxsize=60mm \epsfbox{.eps}
%\end{center}
\caption{ Example of the metastability area
 (between the curves) of the vortex-like state. The upper curve (filled
 circles) is the boundary of the vortex-like state stability region, the lower curve(open
 circles) is the phase boundary between the vortex-like and
single-domain ground states. }
\label{metastability}
\end{figure}

But the phase boundary between the
two ground states is not important because at the large area
around this curve on the phase diagram both the single domain and
the vortex-like states are stable. When we cross the phase boundary
from a large size to a smaller one, the vortex-like state becomes
metastable, i.e., it is not the ground state but it does not
transform to the single-domain state. This is also true for the
single-domain state. The example of the area of metastability is
shown on Fig.\ref{metastability}.

\begin{figure}[h]
 \centerline{\includegraphics[width=0.7\linewidth]{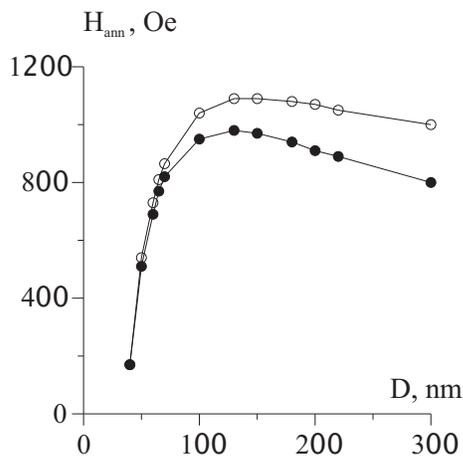}}
 %\begin{center}
 %\epsfxsize=70mm \epsfbox{.eps}
 %\end{center}
 \caption{ The critical field $H_{ann}$ of
 vortex annihilation as a function of diameter $D$. The
 height of the particle $h=20 nm$. The curve marked with filled circles is for
 $\lambda_L=\infty$, with open circles- for $\lambda=50
 nm$.}
 \label{ann}
 \end{figure}

\begin{figure}[h]
 \centerline{\includegraphics[width=0.7\linewidth]{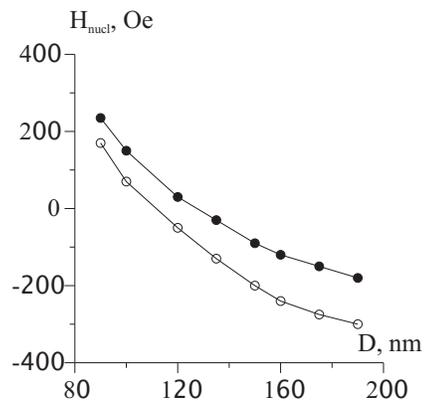}}
 %\begin{center}
 %\epsfxsize=70mm \epsfbox{.eps}
 %\end{center}
 \caption{ The critical field $H_{nucl}$ of vortex nucleation
 as a function of diameter $D$. The
 height of the particle $h=20 nm$. The curve marked with filled circles is for
 $\lambda_L=\infty$, with empty circles- for $\lambda=50
 nm$.}
 \label{nucl}
 \end{figure}

Thus it is more important to investigate the boundary between the
area of metastability and the absolute instability of the
vortex-like or the single-domain
 state. This boundary depends on the external magnetic field
 ${\bf H}_0$. It was found out that applying the in-plane field ${\bf H}_0$
we can cause nucleation of the vortex, i.e., make the
single-domain state unstable, annihilation of
the vortex, i.e. make the vortex-like
state unstable\cite{Guslienko,Frayerman_phr}. Practically, it is more
convenient to vary the external magnetic field than the dimensions
of the particle. That is why we investigate the vortex annihilation,
$H_{ann}$ (Fig.\ref{ann}) and nucleation, $H_{nucl}$
(Fig.\ref{nucl}) fields as functions of particle diameter $D$.
The height of the particle is fixed: $h=20 nm$.

\begin{figure}[h]
  \centerline{\includegraphics[width=1.0\linewidth]{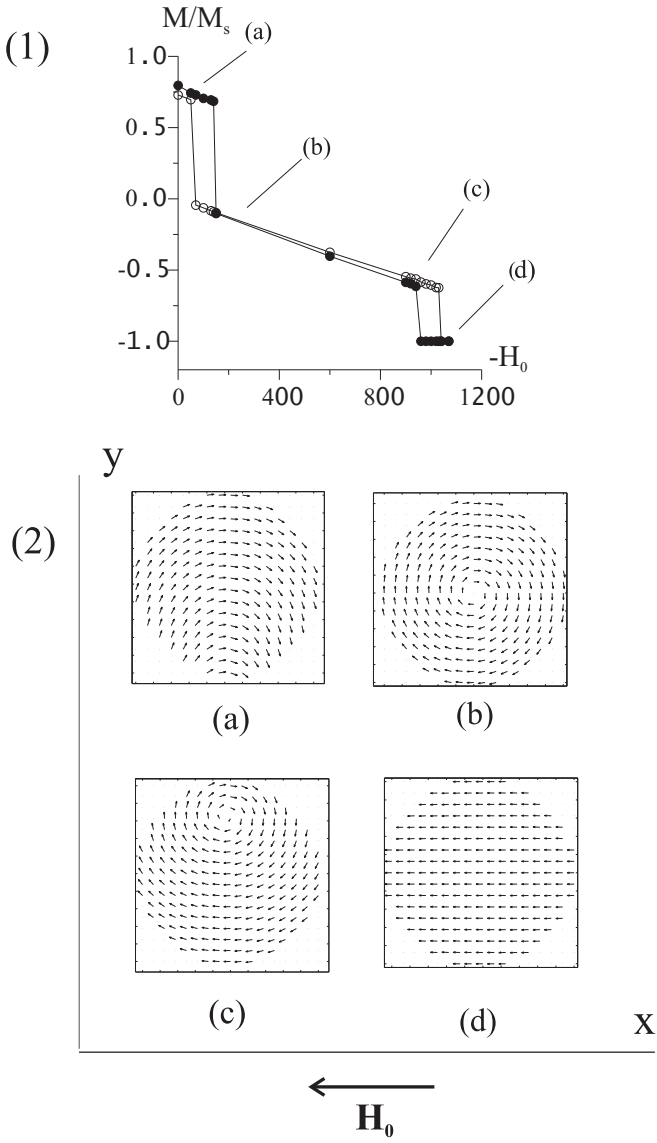}}
 %\begin{center}
 %\epsfxsize=60mm \epsfbox{.eps}
 %\end{center}
 \caption{ (1) The process of magnetization reversal for
 $\lambda_L=\infty$ (open circles) and $\lambda_L=50 nm$ (filled
 circles). The particle dimensions are $h=20 nm$, $D=100 nm$,
 (a),(d)-single-domain states, (b),(c)-vortex-like state. (2) The
 evolution of magnetization distribution during the process of
 magnetization reversal.}
 \label{magn_curve+}
 \end{figure}

 The nucleation and annihilation of the vortex are parts of
 the process of magnetization reversal in a particle driven by
 the external magnetic field. If the particle is initially found
 in the single-domain state, then by applying external field ${\bf
 H}_0$ in the opposite to the magnetic moment direction we
 first stimulate a transition to the vortex-like state, i.e.,
 nucleation of the vortex. When we increase the external field further,
 the vortex annihilates and the particle comes to the
 single-domain state again but with a reversed magnetic moment. Note
 that the annihilation field $H_{ann}$ as always positive for the
 particle with $h=20 nm$ (see Fig.\ref{ann}). The nucleation
 field $H_{nucl}$ becomes negative when the diameter $D$ is
 large enough(see Fig.\ref{nucl}). This means
 that the single domain state is unstable and the external field should
 be applied in order to prevent the nucleation of the vortex-like state.
 In more detail we investigate the
 magnetization reversal for a particle of $D=100 nm$ and
 height $h=20 nm$. At the zero field the particle is in the single-domain state
 with an average magnetic moment directed along the $x$ axis. The
 external field is applied in the opposite direction to the
 initial magnetic moment. The magnetization curve for this particle is
 shown on Fig.\ref{magn_curve+}(1). This curve describes the dependence
 of the average $x$-component of the magnetic moment $M_x$ on the external magnetic field.
 The distribution of magnetization at the stages of
 the magnetization reversal process is shown on
 Fig.\ref{magn_curve+}(2).

\begin{figure}[h]
    \centerline{\includegraphics[width=0.7\linewidth]{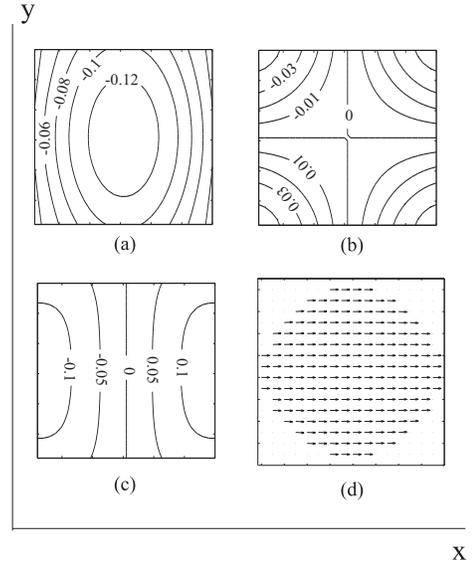}}
% \begin{center}
% \epsfxsize=60mm \epsfbox{.eps}
 %\end{center}
 \caption{(a)-(c) The level curves for the
 components $<H_{sx}>$, $<H_{sy}>$, $<H_{sz}>$ of the magnetic
 field produced by the superconducting current. (d)-magnetization
 of the particle. $D=150 nm$, $h=20 nm$, $\lambda_L= 50 nm$.}
 \label{magn_field}
 \end{figure}

 The influence of a superconductor shows up in the reduced value of the
 average magnetic moment, because it decreases the
 interaction energy $E_{int}$. Therefore, the nucleation field for
 the particle placed above the superconductor is smaller and the
 annihilation field is larger than for the particle in the
 absence of superconductor. The difference between these fields increases
 with the size of the particle(see Fig.\ref{ann}, \ref{nucl}). When the diameter is
 quite large: $D>100 nm$, the shift of the nucleation and
 annihilation fields is about $100-200 oe$.  As we have noted before,
 this shift should be of order of the field produced by
 the superconducting current ${\bf B}_s$. To verify our results we
 find $<{\bf H}_s>$ averaged over the $z$ coordinate:
 $<{\bf H}_s>(x,y)=\int_0^h{\bf H}_s(x,y,z) dz/h$.
 For an example we take a single-domain particle with
 $D= 100 nm$ and $h=20 nm$. On
 Fig.\ref{magn_field}(a),(b),(c) we show the components of $<{\bf H}_s>$
 inside the particle. The magnetic field is normalized to the
 saturation magnetization $M_s=800 Oe$.

 The x-component of magnetic field $<{\bf H}_s>$
 is the largest and it varies from $0.12M_s\approx 100 Oe$ at the center of the particle to
 $0.06M_s\approx 50 Oe$ at the edges. Thus $\Delta H_{nucl}$ and $\Delta
 H_{ann}$ should be about $50\div 100 Oe$.
  According to the results of simulation, the shift of the nucleation field for the cylindrical particle with
 $D=150 nm$ is $\Delta H_{nucl}=110 Oe$(see Fig.\ref{nucl}) and the shift of the annihilation field is
 $\Delta H_{nucl}=-120 Oe$(see Fig.\ref{ann}). Therefore, our
 estimation gives the right order of $\Delta H_{ann}$ and $\Delta
 H_{nucl}$.

 Let us now consider some possible experimental investigations
 based on the effects which we have described. Placing a ferromagnetic
 particle with diameter $D=100-200 nm$ and height $h=20 nm$ above the
 superconductor and cooling it below its critical temperature $T_c$ we will have well observable
 shifts $\Delta H_{nucl}$ and $\Delta H_{ann}$ of about $100-200 Oe$.
 Moreover, the shift of the annihilation field $\Delta H_{ann}$
 enables us to realize a selective magnetization reversal in an array
 of ferromagnetic particles placed above the superconductor. If we neglect the interparticle
 interaction, then the magnetization curve of each particle is
 like the one shown on Fig.\ref{magn_curve+}.

    \begin{figure}[h]
    \centerline{\includegraphics[width=0.9\linewidth]{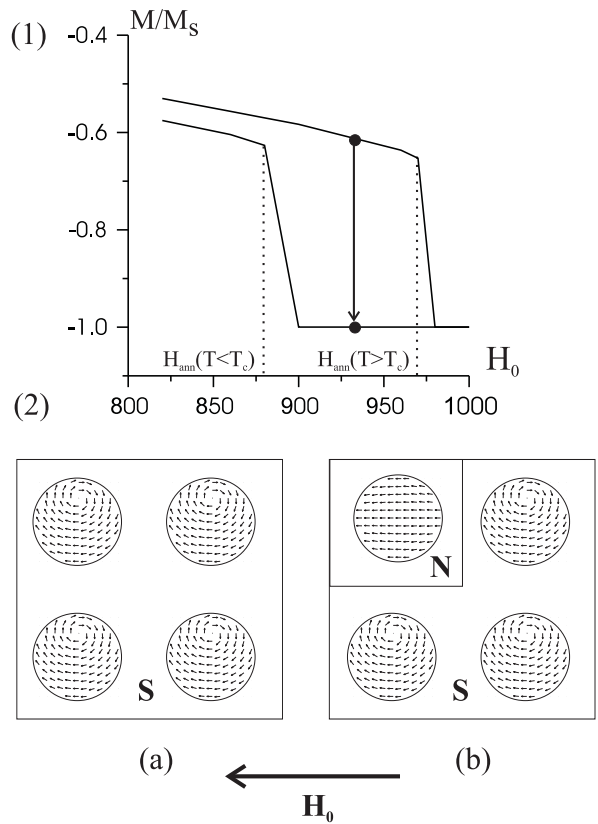}}
 %\begin{center}
 %\epsfxsize=70mm \epsfbox{.eps}
 %\end{center}
 \caption{ (1)-Part of the magnetization curve near
 the annihilation of the vortex. (2)-Selective magnetization reversal in an
 array of ferromagnetic particles.}
 \label{experiment}
 \end{figure}

 Let us assume that initially all particles in the array
 are in the vortex-like state. Applying the magnetic field of
 magnitude between $H_{ann}(T<T_c)$ and $H_{ann}(T>T_c)$(see Fig.\ref{experiment}(1)) will lead
 to a reversible displacement of the vortex core within each
 particle (see Fig.\ref{experiment}(2a)). If then we destroy the
 superconductivity around one of the particles, it will immediately
 fall into the single-domain state (see Fig.\ref{experiment}(2b)).
 By removing the magnetic field we will have all particles
 get back to the initial state except for the one which will remain in the
 single-domain state. Thus we can operate with a single particle in an array
 without disturbing the state of other particles.

  \section{Conclusion} \label{sec:conclusion}
 We presented the results of the numerical investigation of
 the magnetization reversal process in an external magnetic field of a ferromagnetic
 particle placed above the surface of a superconductor. The
 numerical simulation is based on solving the
 Landau-Lifshitz-Gilbert equation for the dynamics of
 magnetic moment. The superconductor is in the Meijssner state and the only
 parameter that affects the interaction between the particle and the superconductor
 is the London penetration depth $\lambda_L$.

 We have shown that for a realistic value of $\lambda_L=50 nm$
 the interaction energy is much smaller than the self-energy of the
 particle and the ground state of the particles does not change
 significantly. But nevertheless, the magnetic field generated by the
 superconducting current
 leads to a decrease of the nucleation field and to an increase of the annihilation
 field by $100-200 Oe$. Based on the effect of the annihilation field shift when the
 superconductor is cooled to temperatures below $T_c$ we describe the method
 of a selective magnetization reversal in an array of ferromagnetic particles.

 \section*{Acknowledgments}

 This work was supported by the Russian Foundation for Basic Research,
 Grant No. 03-02-16774 and Materials Theory Institute at the Argonne
 National Laboratory.

\end{document}